\documentclass[prb,aps,twocolumn,superscriptaddress,10pt,showpacs,showkeys]{revtex4-2}
\usepackage[utf8]{inputenc}
\usepackage{graphicx}
\usepackage{amsmath}
\usepackage{amsfonts}
\usepackage{color}
\usepackage{bm}
\usepackage{tabularx}
\usepackage[separate-uncertainty=true]{siunitx}
\usepackage{natbib}

\begin{document}
\title{A higher-dimensional geometrical approach for the classification of 2D square-triangle-rhombus tilings}
\author{Marianne Imperor-Clerc}
\email{marianne.imperor@universite-paris-saclay.fr}
\affiliation{Laboratoire de Physique des Solides, CNRS and Université Paris-Saclay, 91400 Orsay, France}
\author{Pavel Kalugin}
\affiliation{Laboratoire de Physique des Solides, CNRS and Université Paris-Saclay, 91400 Orsay, France}
\author{Sebastian Schenk}
\affiliation{Institute of Physics, Martin-Luther-Universität Halle-Wittenberg, D-06099 Halle, Germany}
\author{Wolf Widdra}
\affiliation{Institute of Physics, Martin-Luther-Universität Halle-Wittenberg, D-06099 Halle, Germany}
\author{Stefan Förster}
\email{stefan.foerster@physik.uni-halle.de}\affiliation{Institute of Physics, Martin-Luther-Universität Halle-Wittenberg, D-06099 Halle, Germany}
\date{\today}
\begin{abstract}
Square-triangle-rhombus ($\mathcal{STR}$) tilings are encountered in various self-organized multi-component systems. They exhibit a rich structural diversity, encompassing both periodic tilings and long-range ordered quasicrystals, depending on the proportions of the three tiles and their orientation distributions. We derive a general scheme for characterizing $\mathcal{STR}$ tilings based on their lift into a four-dimensional hyperspace. In this approach, the average hyperslope ($2 \times 2$) matrix $\mathcal{H}$ of a patch defines its global composition with four real coefficients: $\mathcal{X}$, $\mathcal{Y}$, $\mathcal{Z}$, and $\mathcal{W}$. The matrix $\mathcal{H}$ can be computed either directly from the area-weighted average of the hyperslopes of individual tiles or indirectly from the border of the patch alone.
The coefficient $\mathcal{W}$ plays a special role as it depends solely on the rhombus tiles and encapsulates a topological charge, which remains invariant upon local reconstructions in the tiling. For instance, a square can transform into a pair of rhombuses with opposite topological charges, giving rise to local modes with five degrees of freedom.
We exemplify this classification scheme for $\mathcal{STR}$ tilings through its application to experimental structures observed in two-dimensional Ba-Ti-O films on metal substrates, demonstrating the hyperslope matrix $\mathcal{H}$ as a precise tool for structural analysis and characterization.
\end{abstract}

\maketitle  

\section{Introduction}
The self-organized growth of multi-component systems often results in the formation of complex structures. In many network-forming systems, such as liquid crystals, polymer blends, or metal-organic frameworks at surfaces \cite{zeng2004,urgel2016,hayashida_polymeric_2007,pasens_interface-driven_2017,liu2019a}, these structures can be described in terms of elementary units with shapes like squares or equilateral triangles. Complex mixtures of these units, occurring under intermediate growth conditions, may give rise to the production of uniform structures exhibiting twelve-fold rotational symmetry, similar to dodecagonal quasicrystalline phases \cite{niizeki1987,gahler_crystallography_1988,yamamoto1996,roth1998}. Meanwhile, dodecagonal structures containing additional \SI{30}{\degree} rhombuses have gained increasing attention over the past decade with the emergence of oxide quasicrystals \cite{forster_quasicrystalline_2013,forster2020,schenk_2d_2022,Li_Ce-Ti-O_2023,Merchan_Sr-Ti-O_2022}, along with columnar liquid quasicrystals \cite{zeng2023}.
\par
For such complex phases, the growth conditions usually control only the stoichiometry of the different types of elementary units (e.g., squares and triangles). However, due to geometrical constraints on packing squares and triangles together, global order emerges in these systems. The link between stoichiometry  and global order can be rationalized by lifting the structure into a higher-dimensional space, a technique developed for studying quasicrystals \cite{imperor-clerc_square-triangle_2021}. 
\par
In this paper, we generalize the results of \cite{imperor-clerc_square-triangle_2021} to square-triangle-rhombus ($\mathcal{STR}$) tilings.  After recalling the hyperspace approach, we develop the characterization of $\mathcal{STR}$ tilings. As we shall see, the addition of rhombuses makes the situation more complex, as the stoichiometry between squares and rhombuses is not fixed by the geometrical constraints.  Indeed, a square can be transformed locally in a pair of rhombuses and vice versa.
Finally, this geometrical approach is applied to the classification of atomic-scale $\mathcal{STR}$ tilings observed in Ba-Ti-O films by scanning tunneling microscopy \cite{schenk_full_2019,wuehrl2022,wuehrl2023}. 
\section{Quasicrystals, approximant phases and hyperslope}
Quasicrystals are highly ordered structures exhibiting rotational symmetries (such as five-, eight-, ten-, or twelve-fold ones) forbidden for a periodic lattice in two or three dimensions. Their structure can be conveniently modeled by aperiodic tilings of the plane or three-dimensional space. Such tilings can be lifted into a higher-dimensional space (hyperspace), where they become a subset of a periodic structure  \cite{gahler_crystallography_1988,henley1991,baake2016}. The real tiling can be recovered as a projection of the lifted tiling onto the so-called physical subspace of the hyperspace. It is also often convenient to consider the projection of the lifted structure onto the linear complement to the physical space (the so-called "internal" or "perpendicular" space). It is noteworthy that for all known quasicrystals, the dimension of the internal space equals that of the physical one.

Geometrically, the lifted tiling resembles a corrugated surface in the hyperspace. This surface can be interpreted as a plot of a function from the physical space to the internal one (colloquially called the {\em phason} coordinate). The gradient of this function (which we shall refer to as the {\em hyperslope}) is an important characteristic of the tiling. On the scale of individual tiles, the hyperslope is constant within each tile. On a larger scale, one can speak of the average hyperslope of a tiling patch, or even of the entire tiling. For instance, the hyperslope of an infinite perfect quasicrystalline tiling is zero (since the projection of such a tiling onto the internal space is bounded). Thus, a non-zero average hyperslope can be used as a measure of deviation from the perfect quasicrystalline state. In this context, the hyperslope is often also referred to as {\em phason strain}. This quantity complements the more traditional characteristics of tilings, such as the statistics of tiles of different shapes or orientations, or the occurrences of different local environments \cite{ishimasa2015}.

The average hyperslope of a tiling is well defined only for tilings that obey certain global uniformity conditions \cite{imperor-clerc_square-triangle_2021}. An important case of globally uniform tilings is that of the so-called approximant phases. These are periodic structures, typically with a small hyperslope, which can thus be considered approximations to the aperiodic structure of a perfect quasicrystal. The hyperslope of an approximant phase is entirely defined by its unit cell.

Perfect quasicrystals are not the only examples of structures with long-range order exhibiting forbidden symmetries. It has been argued that such symmetry can arise asymptotically in the limit of large systems and be stabilized by structural entropy rather than the energy of interatomic interactions \cite{henley1990random, henley1991}. It is commonly believed that for such random tiling models, the entropy density is a quadratic function of the local hyperslope \cite{henley1991}. This ``hydrodynamic'' description predicts that the phason coordinate of a random tiling remains bounded in three dimensions but exhibits divergent fluctuations in one and two dimensions (as the square root or as the logarithm of the distance, respectively) \cite{henley1991, cockayne2000}.
\section{Characterization of $\mathcal{STR}$ tilings}
\label{sec:tilings}
\begin{figure}[b]
    \centering
    \includegraphics[width=0.45\textwidth]{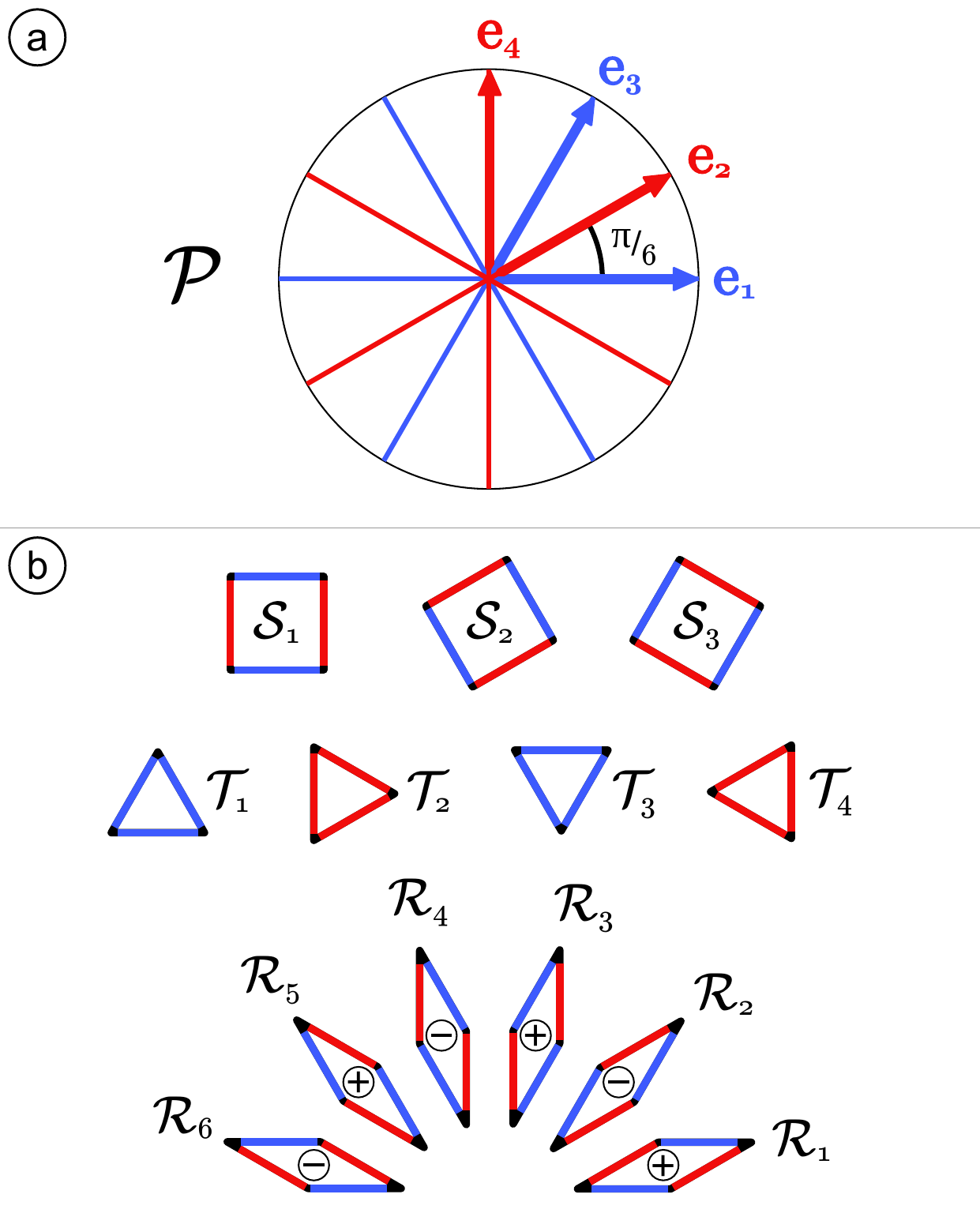}
        \caption{(a) Definition of four vectors in plane $\mathcal{P}$ used for indexing the vertices of a $\mathcal{STR}$ tiling. (b) The complete set of prototiles in $\mathcal{STR}$ tilings including their edge orientation. The plus and minus signs for rhombuses correspond to their topological charge.}
      \label{fig:ParSpaceTiles}
\end{figure}

A general way to characterise $\mathcal{STR}$ tilings is by using the area fraction of each type of tiles. Indeed, this description is independent of the total area of a finite patch and is still well-defined for infinite tilings. 
In $\mathcal{STR}$ tilings all tiles have a common edge length and all edges are aligned in multiples of \SI{30}{\degree}. Accordingly, these tiles can be characterised by using four unit vectors ($\mathbf{e}_1,\mathbf{e}_2,\mathbf{e}_3,\mathbf{e}_4$) and their linear combinations in the physical plane $\mathcal{P}$ as depicted in Fig. \ref{fig:ParSpaceTiles}(a).

Each tile of the tiling is thus a translated copy of one of the 13 prototiles shown in Fig. \ref{fig:ParSpaceTiles}(b). For convenience, we assume one edge to be parallel to $\mathbf{e}_1$. These edges as well as those obtained by the rotation by $\pm \pi/3$ (i.e. those parallel to $\mathbf{e}_3$ and $\mathbf{e}_3-\mathbf{e}_1$) are colored in blue, while all other edges are shown in red. In total, we identify 13 different prototiles in $\mathcal{STR}$ tilings by including their orientations. Squares can occur in three different orientations ($\mathcal{S}_1$, $\mathcal{S}_2$ and $\mathcal{S}_3$), triangles in four ($\mathcal{T}_1$, $\mathcal{T}_2$, $\mathcal{T}_3$ and $\mathcal{T}_4$), and rhombuses in six ($\mathcal{R}_1$ to $\mathcal{R}_6$). In Fig. \ref{fig:ParSpaceTiles}(b), the plus and minus signs correspond to the topological charge associated to the rhombuses, as discussed later on in section \ref{sec:topological_charge}. 
\par
The composition of a $\mathcal{STR}$ tiling is consequently defined by a set of 13 area fractions denotes as $\sigma_i$ for squares, $\tau_i$ for triangles and $\rho_i$ for rhombuses. Indices refer to the orientation of a tile as drawn in Fig.~\ref{fig:ParSpaceTiles}(b). For any finite patch or any infinite tiling without holes or overlaps of tiles, the coverage relation states that the sum of all 13 area fractions is equal to one:
\begin{equation}
    \sigma+\tau+\rho=1 
    \label{eq:coverage}
\end{equation}
with
\begin{equation}
\begin{aligned}
    \sigma&=\sigma_{1}+\sigma_{2}+\sigma_{3}\\
    \tau&=\tau_{1}+\tau_{2}+\tau_{3}+\tau_{4}\\
    \rho&=\rho_{1}+\rho_{2}+\rho_{3}+\rho_{4}+\rho_{5}+\rho_{6}
\end{aligned}
\end{equation}
It is also convenient to introduce $\rho_{+}$ and $\rho_{-}$, the area fractions of rhombuses carrying the same topological charge (see section \ref{sec:topological_charge}):
\begin{equation}
\begin{aligned}
    \rho&=\rho_{+}+\rho_{-}\\
    \rho_{+}&=\rho_{1}+\rho_{3}+\rho_{5}\\
    \rho_{-}&=\rho_{2}+\rho_{4}+\rho_{6}
\end{aligned}
\end{equation}
\par
As a simple example, the area fraction occupied by squares in a patch containing $N_S$ squares is defined as $\sigma=N_S\mathcal{A}_S/(N_T\mathcal{A}_T+N_S\mathcal{A}_S+N_R\mathcal{A}_R)$ where $N_T$ is the number of triangles, $N_R$ the number of rhombuses and the tile's areas are $\mathcal{A}_{S}=a^2$, $\mathcal{A}_T=a^2\sqrt{3}/4$ and $\mathcal{A}_{R}=a^2/2$.   

\subsection{Lifted tiles and their hyperslope}
\label{sec:hyperspace}
\begin{figure}
    \centering
    \includegraphics[width=0.45\textwidth]{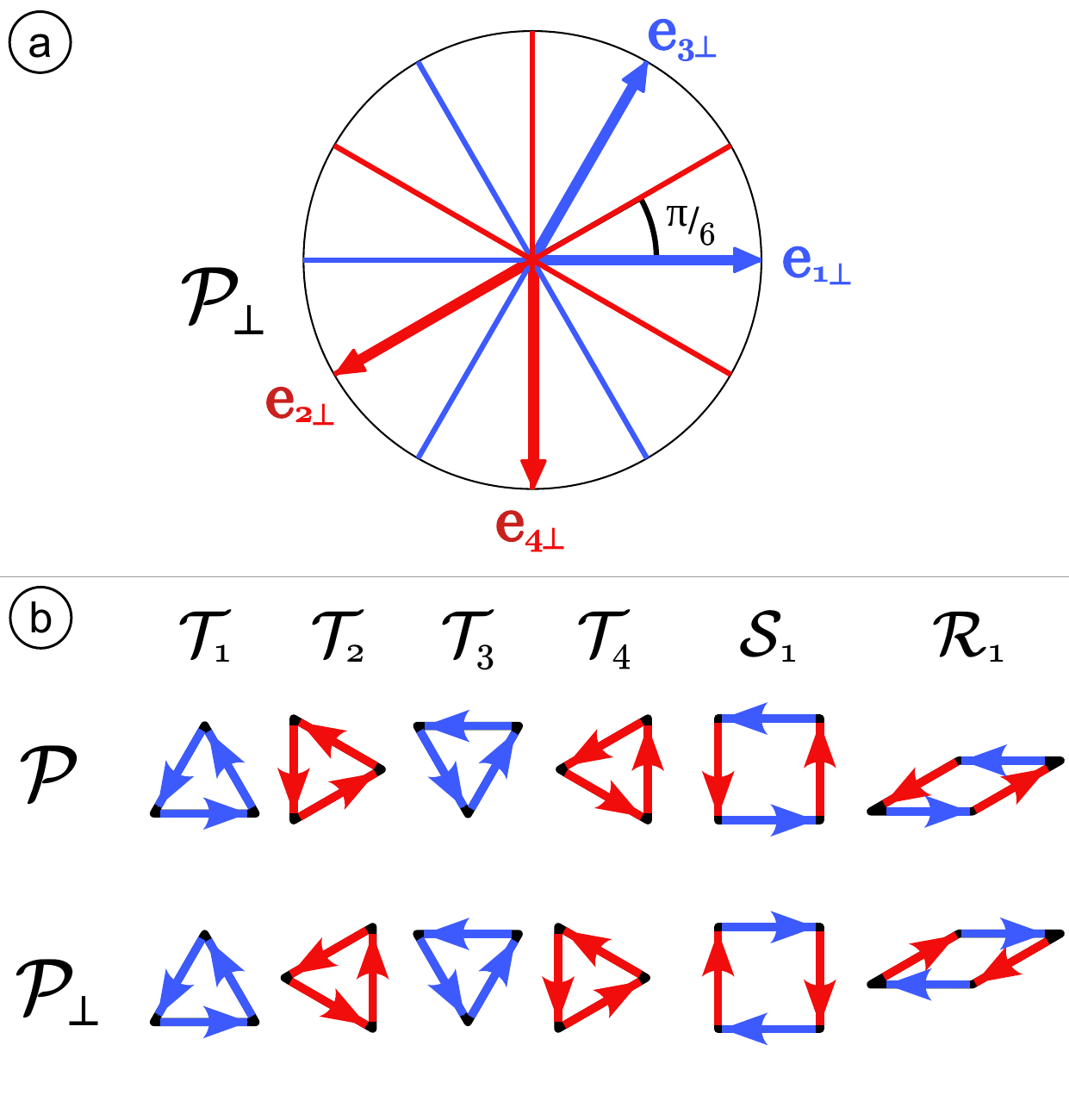}
        \caption{(a) Perpendicular space projection (b) Comparison of the algebraic areas of the tiles in plane $\mathcal{P}$ and in the perpendicular plane $\mathcal{P}_{\bot}$.}
      \label{fig:PerpSpaceTiles}
\end{figure}

Since periodic lattices compatible with the symmetry of a regular dodecagon first appear in four dimensions, it is natural to consider the lifting of an $\mathcal{STR}$ tiling to four-dimensional hyperspace. The lifting procedure is quite similar to the case of square-triangle tilings, described in detail in \cite{imperor-clerc_square-triangle_2021}. It begins with the observation that the coordinates of vertices in the tiling (relative to an arbitrarily chosen reference vertex) are integer linear combinations of four vectors $\{\mathbf{e}_1, \mathbf{e}_2, \mathbf{e}_3, \mathbf{e}_4\}$ shown in Fig.\ref{fig:ParSpaceTiles}(a),where $a$ is the edge length of the tiles.
\begin{equation}
\label{eq:par_basis}
\mathbf{e}_i=a\begin{pmatrix}
\cos\frac{\pi(i-1)}{6}\\ \sin\frac{\pi(i-1)}{6}
\end{pmatrix}.
\end{equation}
Thus, we can index all tile vertices by the coefficients of this combination, a tuple $n=(n_1,n_2,n_3,n_4)$, and associate with a vertex $\mathbf{v}$ its lifted counterpart, a point $\mathbf{V}$ of the four-dimensional periodic lattice:
\begin{equation}
\begin{aligned}
\mathbf{v}&=n_1\mathbf{e}_1+n_2\mathbf{e}_2+n_3\mathbf{e}_3+n_4\mathbf{e}_4\\
\mathbf{V}=\begin{pmatrix}\mathbf{v}\\ \mathbf{v}_{_\bot}\end{pmatrix}&=n_1\bm{\epsilon}_1+n_2\bm{\epsilon}_2+n_3\bm{\epsilon}_3+n_4\bm{\epsilon}_4
\end{aligned}
\label{eq:lifting}
\end{equation}
Here, the 4D vectors $\bm{\epsilon}_i$ form a basis of a lattice in the hyperspace. This hyperspace can be conveniently represented as a direct sum $\mathcal{P} \oplus \mathcal{P}_\bot$ of the physical space $\mathcal{P}$ and its perpendicular complement $\mathcal{P}_\bot$. The original vertex $\mathbf{v}$ is then the projection of $\mathbf{V}$ onto $\mathcal{P}$. Similarly, the projection of $\mathbf{V}$ onto $\mathcal{P}_\bot$ yields the ``perpendicular'' vertex $\mathbf{v}_{_\bot} \in \mathcal{P}_\bot$:
$$
\mathbf{v_{_\bot}}=n_1{\mathbf{e}_1}_{_\bot}+n_2{\mathbf{e}_2}_{_\bot}+n_3{\mathbf{e}_3}_{_\bot}+n_4{\mathbf{e}_4}_{_\bot},
$$
where the vectors ${\mathbf{e}_i}_{_\bot} \in \mathcal{P}_\bot$ (Fig. \ref{fig:PerpSpaceTiles}(a)) are the projections of the basis vectors of the lattice $\bm{\epsilon}_i$ onto $\mathcal{P}_\bot$:
\begin{equation}
{\mathbf{e}_i}_{_\bot}=a\begin{pmatrix}
\cos\frac{7\pi(i-1)}{6}\\ \sin\frac{7\pi(i-1)}{6}
\end{pmatrix}
    \label{eq:perp_basis}
\end{equation} 
\par
In addition to vertices, tiles also have their counterparts in hyperspace. Note that the vertices of a tile remain coplanar after lifting. This fact is trivial for triangular tiles and can be easily verified for squares and rhombuses. Indeed, due to the linearity of equation (\ref{eq:lifting}), four vertices forming a parallelogram in $\mathcal{P}$ still form a parallelogram after lifting and thus belong to the same affine 2D plane $\mathcal{D}$ in hyperspace. This plane can be interpreted as the plot of an affine function $\mathcal{P} \to \mathcal{P}_\bot$. Such a function can be naturally expressed in the following form:
\begin{equation}  
    \begin{pmatrix}
    x_{\bot} \\ y_{\bot} 
    \end{pmatrix}=\mathcal{H} \begin{pmatrix}
    x \\ y 
    \end{pmatrix}+\text{const},
    \label{eq:h_matrix}
\end{equation}
where the $2 \times 2$ matrix $\mathcal{H}$ is called the {\em hyperslope} of the corresponding tile. Note that all tiles of the same shape and orientation have identical hyperslopes. Conversely, with the exception of two pairs of triangular tiles, tiles of different species have different hyperslopes (see Table \ref{tab:hyperslopes}). 

It is often convenient to write the matrix (\ref{eq:h_matrix}) in the following form (see Appendix A):
\begin{equation}  
    \mathcal{H}= \begin{pmatrix}
    \mathcal{Z}+\mathcal{X} & -\mathcal{Y}-\mathcal{W} \\-\mathcal{Y}+\mathcal{W} & \mathcal{Z}-\mathcal{X}
    \end{pmatrix},
    \label{eq:hyperslope_XYZW}
\end{equation}
where $\mathcal{X}$, $\mathcal{Y}$, $\mathcal{Z},$ and $\mathcal{W}$ are independent real coefficients. It is noteworthy that the coefficient $\mathcal{W}$ is zero for all tiles except rhombuses. Thus, the hyperslope matrix of square and triangular tiles is symmetric, while for rhombuses it contains an antisymmetric part $\pm A^R$ (see Table \ref{tab:hyperslopes}):
\begin{equation}
A^{R}=\begin{pmatrix} 0 & -\sqrt{3} \\\sqrt{3} & 0 \end{pmatrix}
\label{eq:A_R}
\end{equation}   
As a matter of fact, the important difference of $\mathcal{STR}$ tilings with respect to $\mathcal{ST}$ tilings \cite{imperor-clerc_square-triangle_2021} is that the presence of rhombuses adds an antisymmetric component to the hyperslope with the coefficient $\mathcal{W}$.
\begin{table}
	\begin{center}
		\caption{Hyperslopes of the different prototiles. }
		\label{tab:hyperslopes}
		\begin{tabular}{|c|c|}
			\hline
            tile & $\mathcal{H}$\\
            \hline\hline
            $\mathcal{T}_1$ and $\mathcal{T}_3$ & $I_2=\begin{pmatrix} 1 & 0 \\0 & 1 \end{pmatrix}$\\
            \hline
            $\mathcal{T}_2$ and $\mathcal{T}_4$ & $-I_2$\\
            \hline
            $\mathcal{S}_1$ & $B^{S1}=\begin{pmatrix} 1 & 0 \\0 & -1 \end{pmatrix}$ \\
            \hline
            $\mathcal{S}_2$ & $B^{S2}=\begin{pmatrix} -1/2 & -\sqrt{3}/2 \\-\sqrt{3}/2 & 1/2 \end{pmatrix}$\\
            \hline
            $\mathcal{S}_3$ & $B^{S3}=\begin{pmatrix} -1/2 & \sqrt{3}/2 \\\sqrt{3}/2 & 1/2 \end{pmatrix}$\\
			\hline
            $\mathcal{R}_1$ & $A^{R1}=\begin{pmatrix} 1 & -2\sqrt{3} \\0 & -1 \end{pmatrix}=-2B^{S3}+A^{R}$\\
			\hline
            $\mathcal{R}_2$ & $A^{R2}=\begin{pmatrix} -2 & \sqrt{3} \\-\sqrt{3} & 2 \end{pmatrix}=-2B^{S1}-A^{R}$\\
			\hline
            $\mathcal{R}_3$ & $A^{R3}=\begin{pmatrix} 1 & 0 \\2\sqrt{3} & -1 \end{pmatrix}=-2B^{S2}+A^{R}$\\
			\hline
            $\mathcal{R}_4$ & $A^{R4}=\begin{pmatrix} 1 & 0 \\-2\sqrt{3} & -1 \end{pmatrix}=-2B^{S3}-A^{R}$\\
			\hline
            $\mathcal{R}_5$ & $A^{R5}=\begin{pmatrix} -2 & -\sqrt{3} \\\sqrt{3} & 2 \end{pmatrix}=-2B^{S1}+A^{R}$\\
			\hline
            $\mathcal{R}_6$ & $A^{R6}=\begin{pmatrix} 1 & 2\sqrt{3} \\0 & -1 \end{pmatrix}=-2B^{S2}-A^{R}$\\
            \hline
            anti-symmetric part* & $A^{R}=\begin{pmatrix} 0 & -\sqrt{3} \\\sqrt{3} & 0 \end{pmatrix}$\\
			\hline
		\end{tabular}
	\end{center}
\end{table}

Let's now compare the projections of lifted tiles onto the physical and perpendicular spaces. For this purpose, it is convenient to distinguish two types of edges, shown in red and blue colors in Figure \ref{fig:ParSpaceTiles}. As inferred from equations (\ref{eq:par_basis}) and (\ref{eq:perp_basis}), the blue edges have the same direction in $\mathcal{P}$ and $\mathcal{P}_\bot$, while the directions of the red edges in the two spaces are opposite (see Figure \ref{fig:PerpSpaceTiles}a). Consequently, the two projections of a lifted tile appear exactly the same except for the orientation of the edge vectors. It's worth noting that $\mathcal{T}_2$ and $\mathcal{T}_4$ are swapped in the two projections. 

Although the projections of lifted tiles onto $\mathcal{P}_{\bot}$ have the same shape as the original tiles in the physical space $\mathcal{P}$, the orientation of their edges is not preserved (see Figure \ref{fig:PerpSpaceTiles}b). The counterclockwise orientation of the edges in the physical space remains counterclockwise in $\mathcal{P}_{\bot}$ for triangles, whereas for squares and rhombuses, it becomes clockwise, due to the inversion of signs for the red vectors. This fact can also be derived from equation (\ref{eq:h_matrix}), as the ratio by which the algebraic area is altered by the affine map (\ref{eq:h_matrix}) equals the determinant of the hyperslope $\mathcal{H}$. Indeed, the determinant of the hyperslope for triangular tiles is $1$, whereas for squares and rhombuses, it equals $-1$ (see Table \ref{tab:hyperslopes}).

\subsection{Finite patch: projected areas and average hyperslope}
\label{sec:average_hyperslope}
Let's consider a finite simply connected patch of $\mathcal{STR}$ tiling, and its lifted counterpart, obtained by lifting all of its tiles (section \ref{sec:hyperspace}). In this section, we shall be interested in average characteristics of such a patch. Let us start by comparing the areas of its projections on $\mathcal{P}$ and $\mathcal{P}_{\bot}$. In plane $\mathcal{P}$, the algebraic area $\mathcal{A}$ is positive, as it is simply the sum of individual areas of all tiles in the patch. But for the projection on plane $\mathcal{P}_{\bot}$, square and rhombus areas are counted with a negative sign, whereas triangles are counted with a positive sign (section \ref{sec:hyperspace}). As a result, the algebraic area $\mathcal{A}_{\bot}$ is the sum of the areas of triangles minus the sum of the areas of squares and rhombuses. The two algebraic areas of the patch projected on $\mathcal{P}$ and $\mathcal{P}_{\bot}$ are: 
\begin{align*}
    \mathcal{A}&=N_{T}\mathcal{A}_{T}+N_{S}\mathcal{A}_{S}+N_{R}\mathcal{A}_{R}\\
    \mathcal{A}_{\bot}&=N_{T}\mathcal{A}_{T}-N_{S}\mathcal{A}_{S}-N_{R}\mathcal{A}_{R}
\end{align*}
The ratio of these two areas depends only on the triangle area fraction $\tau$:
\begin{equation}
\frac{\mathcal{A}_{\bot}}{\mathcal{A}} =\tau-(\sigma+\rho)=2\tau-1
\label{eq:area_ratio}
\end{equation}
\begin{figure}
    \centering
    \includegraphics[width=0.45\textwidth]{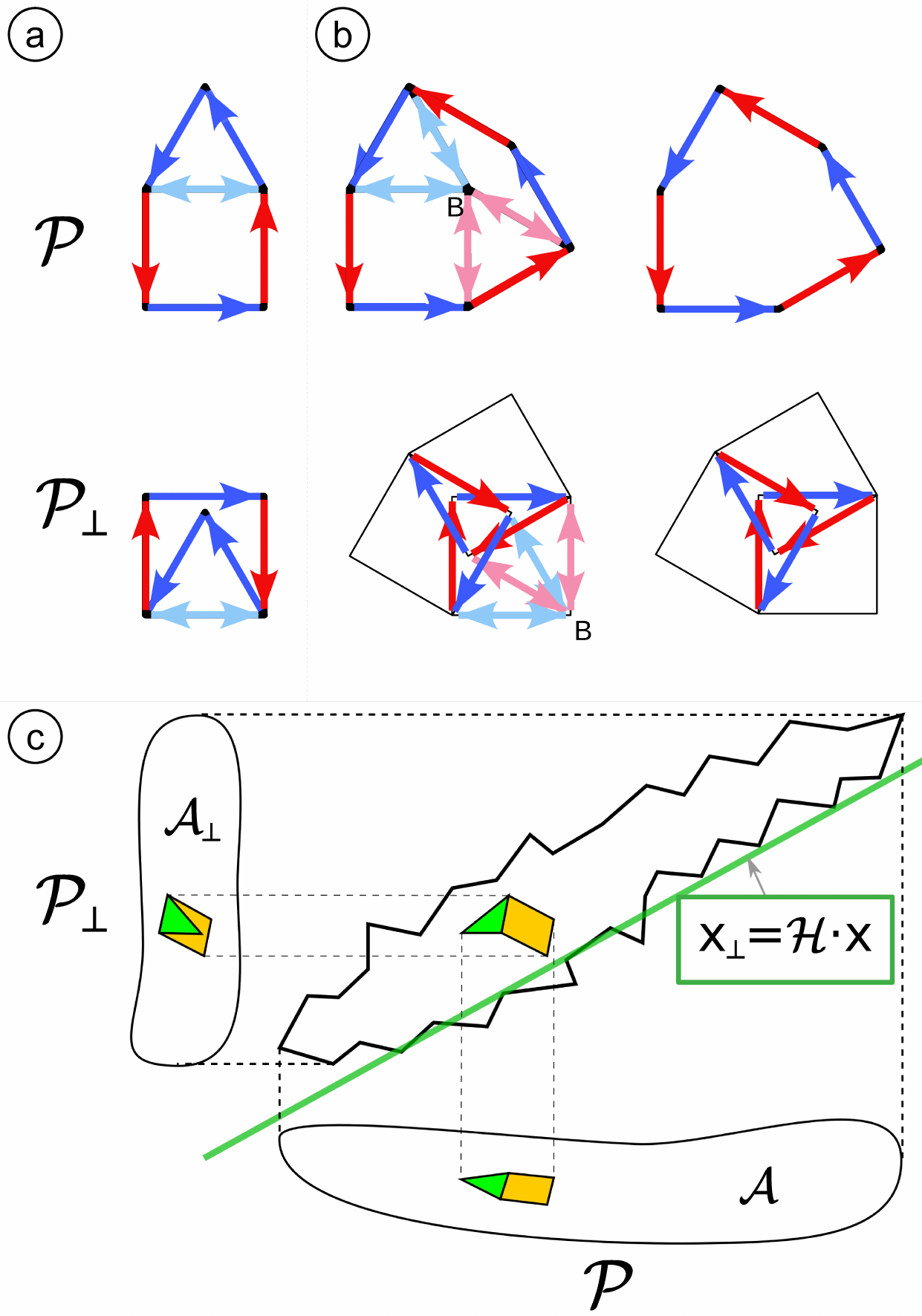}
    \label{fig:lifted-patch}
    \caption{Lifted patch in hyperspace and its two projections on planes $\mathcal{P}$ and $\mathcal{P}_{\bot}$. Edge vectors along the borders are indicated by arrows, while double arrows correspond to the internal edges of the patch. Note that in projection to $\mathcal{P}_\bot$ the internal edges extend beyond the patch border. (a) Example of a house shape consisting of a square and a triangle.  (b) Example of a shield shape with a positive topological charge. Its projection in $\mathcal{P}_{\bot}$ is a three-arm star shape. (c) Illustration of the lifted patch in the hyperspace. The border of the lifted patch (black lines) follows and average plane (green line) $\mathcal{P}_\mathcal{H}$ of hyperslope $\mathcal{H}$.}
\end{figure}
Let us illustrate the above by two simple examples of patches (see Figure \ref{fig:lifted-patch}): a ``house'' made of one square and one triangle (Fig. \ref{fig:lifted-patch}(a)) and a ``shield'' (Fig. \ref{fig:lifted-patch}(b)). Note that in both cases, the algebraic area $\mathcal{A}_{\bot}$ is negative, and its absolute value is smaller than $\mathcal{A}$. Starting from the patch in $\mathcal{P}$, the border of the projected patch in $\mathcal{P}_{\bot}$ is constructed from the sequence of the projected edge vectors. The resulting shape for a ``house'' also contains only two tiles, but its algebraic area $\mathcal{A}_{\bot}$ is negative and equals the area of a triangle minus the area of a square. Indeed, lifting a patch of the tiling and then projecting it onto the perpendicular plane $\mathcal{P}_\bot$ can be visualized as folding a sheet of paper (Figure \ref{fig:lifted-patch}(c)). In this process, all triangles remain on their recto side, while squares and rhombuses are folded to their verso side. For larger patches, the folding will result in the superposition of multiple layers of paper. This is the case for the "shield" shape (Figure \ref{fig:lifted-patch}(b)), which has the following algebraic areas in two projections:
\begin{equation}
\label{eq:shield_areas}
\begin{aligned}
    \mathcal{A}_{\text{shield}}&=\frac{a^2}{2}(3+\sqrt{3})\\
    \mathcal{A}_{\text{shield}\bot}&=-\frac{a^2}{2}(3-\sqrt{3})
\end{aligned}
\end{equation}  

Let us now consider the average hyperslope of a patch, that is, the hyperslope of the 2D plane in $\mathcal{P} \oplus \mathcal{P}_{\bot}$ which fits the most closely the lifted patch (see Fig. \ref{fig:lifted-patch}(b)).
Formally, the average hyperslope is defined as the area-weighted average of hyperslopes of all tiles in the patch:
\begin{equation}
    \begin{split}
    \mathcal{H} &= 
    \sigma_1 B^{S1}+
    \sigma_2 B^{S2}+
    \sigma_3 B^{S3}\\\
    &+\tau_1 I_{1}+
    \tau_2 I_{2}+
    \tau_3 I_{1}+
    \tau_4 I_{2}\\\
    &+\rho_1 A^{R1}+\rho_2 A^{R2}+\rho_3 A^{R3}+\rho_4 A^{R4}+\rho_5 A^{R5}+\rho_6 A^{R6}
    \label{eq:average_hyperslope}
    \end{split}
\end{equation}
Given hyperslopes for the 13 tile species (see Table \ref{tab:hyperslopes}), it is rather straightforward to calculate the four coefficients of the average hyperslope matrix $\mathcal{H}$ from (\ref{eq:average_hyperslope}). Noteworthy, there exists an alternative and more efficient algorithm to compute the average hyperslope only from the boundary of the patch as detailed in Appendix \ref{sec:appendix_Geometry_4D}, that leads to the expressions (\ref{eq:border_hyperslope}) and (\ref{eq:border_hyperslope_det}). This fact is remarkable, since the average hyperslope of a patch is completely determined by its boundary! One of the consequence of this property is that the composition of the patch within a fixed boundary cannot be modified arbitrarily (see more details in section \ref{sec:local_modes}).
\par
Let us now consider the parameterization (\ref{eq:hyperslope_XYZW}) of the average hyperslope by four coefficients ($\mathcal{X}$, $\mathcal{Y}$, $\mathcal{Z}$, $\mathcal{W})$. A remarkable fact is that the coefficient $\mathcal{Z}$ depends solely on the triangles:
\begin{equation}
    \mathcal{Z} = \tau_{1}+\tau_{3}-(\tau_{2}+\tau_{4})
    \label{eq:Z}
\end{equation}
when the coefficient $\mathcal{W}$ depends only on the rhombuses:
\begin{equation}
    \mathcal{W} =\sqrt{3}( \rho_{1}-\rho_{2}+\rho_{3}-\rho_{4}+\rho_{5}-\rho_{6})
    \label{eq:W}
\end{equation}
We are using this fact to define $\mathcal{W}$ as the topological charge of a patch as explained in the next section \ref{sec:topological_charge}.
\par
The two last coefficients ($\mathcal{X}$, $\mathcal{Y}$) of the global composition reflect that the square and rhombuses tiles are intimately coupled. In order to get a more symmetric expression, it may be convenient to replace ($\mathcal{X}$, $\mathcal{Y}$) by a set of three coefficients $(\mathcal{X}_1,\mathcal{X}_2,\mathcal{X}_3)$ with the constraint that $\mathcal{X}_1+\mathcal{X}_2+\mathcal{X}_3 = \sigma-2\rho$:
\begin{equation}
\begin{aligned}
    \mathcal{X} &= \mathcal{X}_1-(\mathcal{X}_2+\mathcal{X}_3)/2\\
    \mathcal{Y} &=\sqrt{3}(\mathcal{X}_2-\mathcal{X}_3)/2
\label{eq:XY}
\end{aligned}
\end{equation}
with 
\begin{equation}
\begin{aligned}
    \mathcal{X}_1 &= \sigma_{1}-2(\rho_{2}+\rho_{5})\\
    \mathcal{X}_2 &= \sigma_{2}-2(\rho_{3}+\rho_{6})\\
    \mathcal{X}_3 &= \sigma_{3}-2(\rho_{4}+\rho_{1})
\label{eq:X1X2X3}
\end{aligned}
\end{equation}
\par
Finally, the average hyperslope matrix can be expressed using five different matrices, $I_2$, $B^{S1}$, $B^{S2}$, $B^{S3}$ and $A^R$ (Tab. \ref{tab:hyperslopes}) 
    \begin{equation}
        \mathcal{H} = 
        \mathcal{X}_1 B^{S1}+
        \mathcal{X}_2 B^{S2}+
        \mathcal{X}_3 B^{S3}+ 
        \mathcal{Z} I_{2}+
        \mathcal{W} A^{R}/\sqrt{3}
 \end{equation} 
yielding 
\begin{equation}
    \mathcal{H} = \begin{pmatrix}
        \mathcal{Z}+\frac{1}{2}(2\mathcal{X}_1-\mathcal{X}_2-\mathcal{X}_3) & -\mathcal{W}+\frac{\sqrt{3}}{2}(\mathcal{X}_3-\mathcal{X}_2) \\ \mathcal{W}+\frac{\sqrt{3}}{2}(\mathcal{X}_3-\mathcal{X}_2) & \mathcal{Z}-\frac{1}{2}(2\mathcal{X}_1-\mathcal{X}_2-\mathcal{X}_3)
    \end{pmatrix}\\
    \label{eq:hyperslope_X1X2X3}
\end{equation}

\subsection{Topological charge}
\label{sec:topological_charge}
Rhombuses are set aside of other tile types since their hyperslope does not possess the so-called ``irrotational property'' \cite{oxborrow_random_1993}. Hence, the circulation of the phason coordinate field over the boundary of a rhombus is non-zero, and the presence of a single rhombus in a patch of a tiling can be detected by analyzing the boundary of the patch only. In this sense, each rhombus is associated with a topological charge, which can be positive (for $\mathcal{R}_1$, $\mathcal{R}_3$ and $\mathcal{R}_5$) or negative (for $\mathcal{R}_2$, $\mathcal{R}_4$ and $\mathcal{R}_6$). Two rhombuses with opposite topological charges and adjacent orientation can recombine together to give a square (see Section \ref{sec:local_modes}), while two rhombuses with the same topological charge cannot recombine at all (see Fig. \ref{fig:shield}).
\par
The coefficient $\mathcal{W}$ in (\ref{eq:hyperslope_XYZW}) represents the antisymmetric part of the average hyperslope of a patch. As follows from (\ref{eq:W}) it is proportional to the density of its net topological charge:
\begin{equation}
    \mathcal{W} =\sqrt{3}(\rho_{+}-\rho_{-})=\sqrt{3}\mathcal{A}_{R}\frac{N_{R+}-N_{R-}}{\mathcal{A}},
    \label{eq:topological_charge}
\end{equation}
where $N_{R+}$ and $N_{R-}$ are the numbers of rhombuses carrying positive and negative topological charge:
\begin{equation*}
\begin{aligned}
N_{R+}&=N_{R_1}+N_{R_3}+N_{R_5}\\
N_{R-}&=N_{R_2}+N_{R_4}+N_{R_6}.
\end{aligned}
\end{equation*}
Thus, the case of zero net topological charge corresponds to $\mathcal{W}=0$ and means that the two subsets of orientations ($\mathcal{R}_1$, $\mathcal{R}_3$, $\mathcal{R}_5$) and ($\mathcal{R}_2$, $\mathcal{R}_4$, $\mathcal{R}_6$) are balanced with equal area fractions ($\rho_{+}=\rho_{-}$). This case is illustrated for example by the patches shown in Figure \ref{fig:dodecagons}. 
\par
A rhombus often appears as a part of the so-called  ``shield'' shape, an irregular hexagon with three-fold symmetry (Figure \ref{fig:lifted-patch}(b)). This shape can be filled with a square, a rhombus, and two triangles. Noteworthy, shield-shaped holes are common defects in real atomic structures modeled by $\mathcal{STR}$ tilings, e.g. they have been observed as atomic vacancies in Ba-Ti-O layers by STM imaging \cite{forster_quasicrystalline_2013,schenk_full_2019}. A shield can appear in two orientations (Figure \ref{fig:lifted-patch}(c)), and depending on the orientation, it carries a positive or negative topological charge. An interesting fact is that the average hyperslope for a shield (oriented as shown in Fig. \ref{fig:lifted-patch}(b)) is a purely antisymmetric matrix: 
\begin{equation}
    \mathcal{H}_{\text{shield}}=\frac{1}{3+\sqrt{3}}A^{R}=\begin{pmatrix}
        0 & -\frac{\sqrt{3}}{3+\sqrt{3}} \\ \frac{\sqrt{3}}{3+\sqrt{3}} & 0
    \end{pmatrix}\\
    \label{eq:hyperslope_shield}
\end{equation}

\subsection{Local modes}
\label{sec:local_modes}
Let us now discuss whether the area fractions occupied by tiles of different shapes and orientations can be changed locally, i.e., within a bounded region, without altering the surrounding tiling. Specifically, we are interested in the constraints on such {\em local modes}. The most obvious constraint arises from the requirement that the tiles must always cover the entire region, leading to the condition $\tau + \sigma + \rho = 1$. Note, however, that the area of a triangle is incommensurate with that of a square or a rhombus. Therefore, a local mode cannot change the total number of triangles within the concerned region. Combined with the first condition, this gives two constraints on the local modification of the tiling composition: $\tau = \mathrm{const}$ and $\sigma + \rho = \mathrm{const}$.

Two additional constraints on the local modes can be obtained by analyzing the dual graph of the tiling. Consider, for instance, the lines of this graph dual to the edges parallel to $\mathbf{e}_1$. If we follow such lines from bottom to top, they can terminate only at the triangles of type $\mathcal{T}_1$, or originate at the triangles of type $\mathcal{T}_3$. Since local modifications within a bounded region cannot change the number of lines of the dual graph entering and exiting it, we must have $\tau_1 - \tau_3 = \mathrm{const}$. Applying the same arguments to the edges of different orientations leads to a second condition: $\tau_2 - \tau_4 = \mathrm{const}$.

The average hyperslope of a bounded region is completely determined by its boundary (see Appendix A) and thus cannot be modified locally. On the other hand, the hyperslope coefficients $\mathcal{X}$, $\mathcal{Y}$, $\mathcal{Z}$, and $\mathcal{W}$ depend only on the area fractions, as follows from equations (\ref{eq:Z}-\ref{eq:X1X2X3}). This results in four additional linear constraints on the modifications of area fractions by local modes.

\begin{figure}
    \centering
    \includegraphics[width=0.45\textwidth]{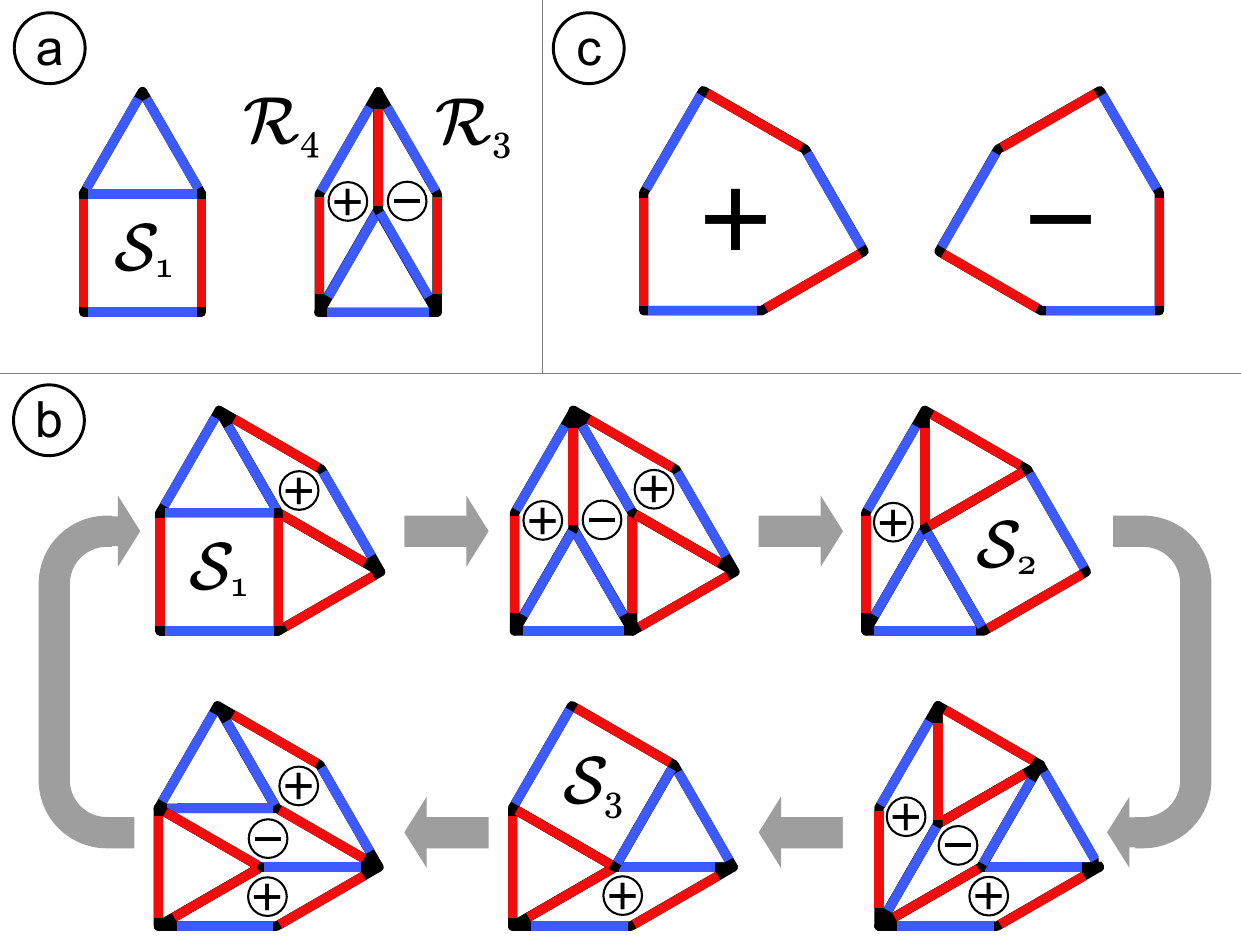}
     \caption{Local modes and topological charge. (a) Local mode inside a ``house'', consisting in the exchange of a square tile by two rhombuses of opposite topological charges. Note that the triangle is translated inside the shape by the local mode. (b) The six possible configurations of tiles inside a ``shield'' with a positive topological charge. A sequence of elementary local transformations (\ref{eq:houses}) allows to obtain all configurations. Note the translations of the two triangles inside the shield from one configuration to the other. (c) Two shields with opposing topological charges.}
      \label{fig:shield}
\end{figure}

We have seen that the area fractions of 13 tile species in an $\mathcal{STR}$ tiling within a finite region are constrained by 8 linear conditions. The remaining five degrees of freedom can only correspond to transformations involving squares and rhombuses. An example of such a local mode is provided by the two different ways to tile the ``house'' shape shown in Fig. \ref{fig:shield}(a). Considering different orientations of squares and rhombuses, this leads to six possible transformations:
\begin{equation}
\begin{aligned}
\mathcal{R}_1 + \mathcal{R}_6\quad \longleftrightarrow &\quad \mathcal{S}_1 & \longleftrightarrow\quad \mathcal{R}_3 + \mathcal{R}_4 \\
\mathcal{R}_4 + \mathcal{R}_5\quad \longleftrightarrow &\quad \mathcal{S}_2 & \longleftrightarrow\quad \mathcal{R}_1 + \mathcal{R}_2 \\
\mathcal{R}_2 + \mathcal{R}_3\quad \longleftrightarrow &\quad \mathcal{S}_3 & \longleftrightarrow\quad \mathcal{R}_5 + \mathcal{R}_6
\end{aligned}
\label{eq:houses}
\end{equation}
These six local transformations are not independent, as the sum of the ``reaction equations'' (\ref{eq:houses}) has identical right- and left-hand sides. One can easily verify that (\ref{eq:houses}) contains no other dependencies, thus accounting for all five remaining local modes.
\par
It is remarkable that any local modification of an $\mathcal{STR}$ tiling can be decomposed into a sequence of elementary local transformations (\ref{eq:houses}). This fact can be illustrated with the example of the so-called ``shield'' shape (Fig. \ref{fig:shield}(c)). The symmetry of this shape allows it to be tiled in different ways. Each configuration of tiles includes either a single rhombus or three rhombuses (Fig. \ref{fig:shield}(b)), and thus shields carry a topological charge; the sign of this charge depends on the shield orientation. As shown in Fig. \ref{fig:shield}(b), all possible configurations of a shield can be transformed into each other using the elementary transformations (\ref{eq:houses}).
\par
\begin{figure}
    \centering
    \includegraphics[width=0.45\textwidth]{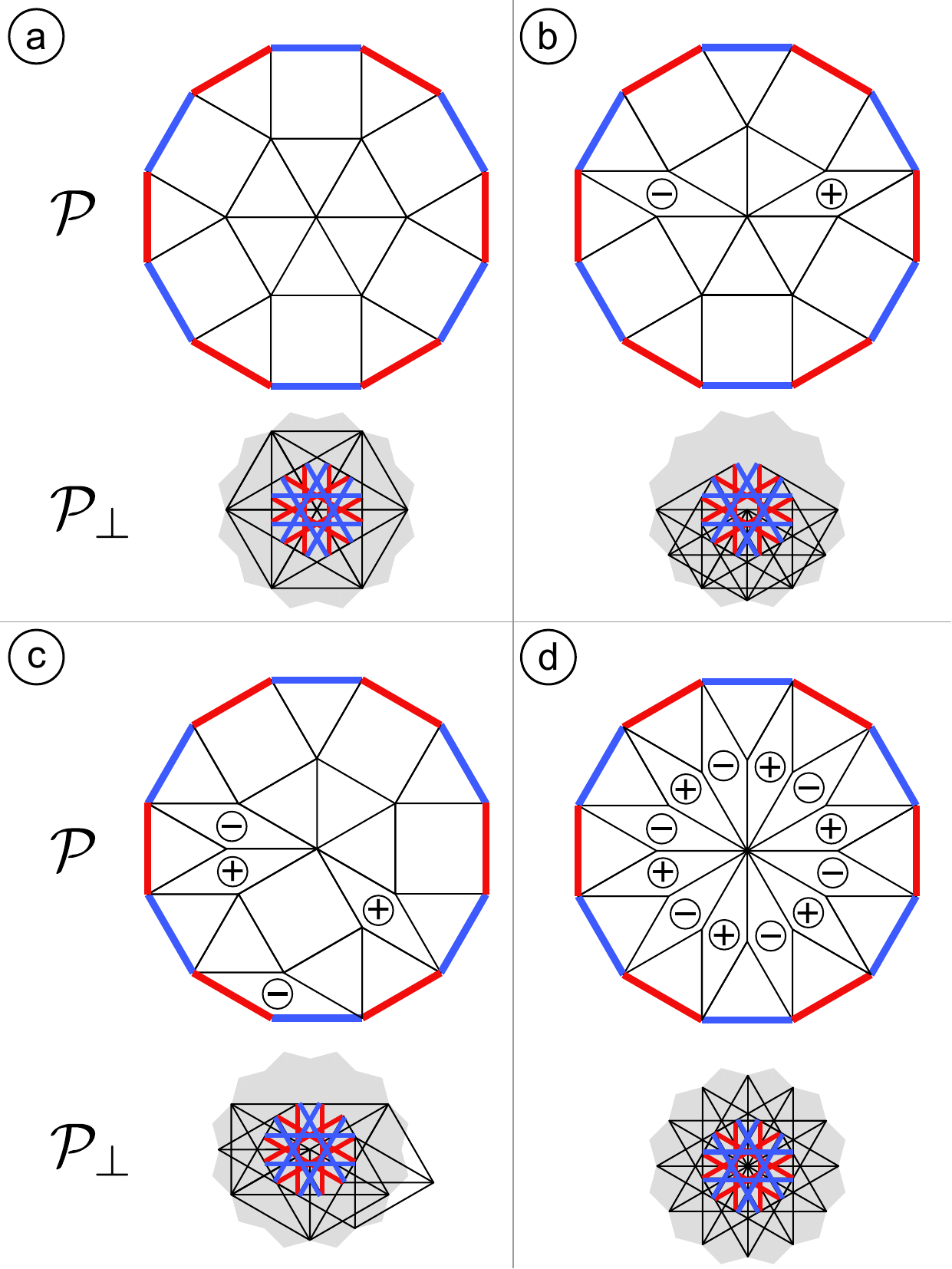}
     \caption{Regular dodecagon filled with four different configurations of tiles. At the bottom are shown the corresponding projections on $\mathcal{P_{\bot}}$. The border of the patch is depicted with a thicker line. The gray shaded area is the acceptance domain for the Niizeki-Gähler quasicrystalline tiling (NGT) \cite{gahler_crystallography_1988}. (a) One of the two configurations without rhombuses ($\sigma=0.536$, $\rho=0$ and $\tau=0.464$). (b) Seed of the NGT ($\sigma=0.446$, $\rho=0.09$ and $\tau=0.464$). Transformation from (a) to (b) replaces a square $\mathcal{S}_1$ by a couple of rhombuses of opposite topological charge. (c) An example of configuration with 2 pairs of rhombuses ($\sigma=0.357$, $\rho=0.179$ and $\tau=0.464$). (d) The most symmetric configuration with no squares and the maximum number of rhombuses ($\sigma=0$, $\rho=0.536$ and $\tau=0.464$).}
      \label{fig:dodecagons}
\end{figure}

Another example of shape which can be filled by squares, triangles and rhombuses in remarkably many ways is the regular dodecagon (see Figure \ref{fig:dodecagons}). It is possible to transform any of these configurations to any other by elementary local transformations (see Fig. \ref{fig:shield}(a)). Thus, different configurations may have different number of squares or rhombuses, but the sum of the corresponding area fractions remains constant: $\sigma+\rho=1-\tau=4-2\sqrt{3}$. Note also that the total topological charge inside the regular dodecagonal patch is always zero.
\par
Among different dodecagonal patches, the ones having a vertex at the center of the dodecagon play a special role of ``seeds'' for inflation-based quasiperiodic tilings. To start with, a possible seed to construct quasicristalline square/triangle tilings is shown in Figure \ref{fig:dodecagons}(a) \cite{imperor-clerc_square-triangle_2021}. The configuration on Figure \ref{fig:dodecagons}(b) is a seed for the Niizeki-Gähler tiling (NGT) (see section \ref{sec:BaTiO}) which is characterised by two rhombuses in orientations $\mathcal{R}_1$ and $\mathcal{R}_6$ sharing only one common vertex \cite{niizeki1987,gahler_crystallography_1988}. Finally, the configuration in Figure \ref{fig:dodecagons}(c) is related to the St\"ampfli tiling, which can be constructed using a grid method with two hexagonal grids \cite{Sadoc_2023}.
\subsection{Rotational symmetry}
\label{sec:symmetry}
Let us now consider the rotation of the $\mathcal{STR}$ tiling  by $\pi/6$ in the physical plane $\mathcal{P}$. Such a rotation permutes the set of prototiles (Figure \ref{fig:ParSpaceTiles}). Note that the permutation is cyclic for the prototiles of the same shape. Therefore, we can apply the lifting procedure to the rotated tiling as well. In particular, this rotation acts on the integer coefficients in (\ref{eq:lifting}) in the following way:
\begin{equation}
    \begin{pmatrix}
        n_1\\n_2\\n_3\\n_4
    \end{pmatrix}
    \mapsto
    \begin{pmatrix}
        0 & 0 & 0 & -1\\
        1 & 0 & 0 & 0\\
        0 & 1 & 0 & 1\\
        0 & 0 & 1 & 0
    \end{pmatrix}
    \begin{pmatrix}
        n_1\\n_2\\n_3\\n_4
    \end{pmatrix}
    \label{eq:rotate_tuple}
\end{equation}
\par
As follows from (\ref{eq:perp_basis}), the counterclockwise rotation by $\pi/6$ in $\mathcal{P}$ corresponds to a clockwise rotation by $5\pi/6$ in the perpendicular space $\mathcal{P}_\bot$. Therefore, we can define the action of this transformation on the hyperslope matrix $\mathcal{H}$ in (\ref{eq:h_matrix}). Using the parameterization (\ref{eq:hyperslope_XYZW}) with four coefficients, this action can be expressed as follows:
\begin{equation}
    \begin{pmatrix}
        \mathcal{X}\\\mathcal{Y}
    \end{pmatrix}
    \mapsto
    \begin{pmatrix}
        -1/2 & -\sqrt{3}/2\\
        \sqrt{3}/2 & -1/2
    \end{pmatrix}
    \begin{pmatrix}
        \mathcal{X}\\\mathcal{Y}
    \end{pmatrix}
    \label{eq:rotation_XY}
\end{equation}
\begin{equation}
    \begin{aligned}
    \mathcal{Z} &\mapsto -\mathcal{Z}\\
    \mathcal{W} &\mapsto -\mathcal{W}
    \end{aligned}
    \label{eq:rotation_ZW}
\end{equation}
Note that (\ref{eq:rotation_XY}) is the rotation by $2\pi/3$ in the $\mathcal{X}$-$\mathcal{Y}$ parameter plane. It can be conveniently expressed as well in terms of the coefficients $\mathcal{X}_1$, $\mathcal{X}_2$, and $\mathcal{X}_3$ in (\ref{eq:XY}):
$$
\mathcal{X}_1 \mapsto \mathcal{X}_2 \mapsto \mathcal{X}_3 \mapsto \mathcal{X}_1
$$
As follows from (\ref{eq:rotation_XY}) and (\ref{eq:rotation_ZW}), the rotation of the physical plane by $\pi$ does not change the hyperslope. Therefore, the full orbit of hyperslopes under the action of the twelve-fold symmetry group contains six elements, with the following coefficients:
\begin{equation}
    \begin{aligned}
    &\mathcal{X}_1,\mathcal{X}_2,\mathcal{X}_3,\mathcal{Z},\mathcal{W}\\
    &\mathcal{X}_2,\mathcal{X}_3,\mathcal{X}_1,-\mathcal{Z},-\mathcal{W}\\
    &\mathcal{X}_3,\mathcal{X}_1,\mathcal{X}_2,\mathcal{Z},\mathcal{W}\\
    &\mathcal{X}_1,\mathcal{X}_2,\mathcal{X}_3,-\mathcal{Z},-\mathcal{W}\\
    &\mathcal{X}_2,\mathcal{X}_3,\mathcal{X}_1,\mathcal{Z},\mathcal{W}\\
    &\mathcal{X}_3,\mathcal{X}_1,\mathcal{X}_2,-\mathcal{Z},-\mathcal{W}
    \end{aligned}
    \label{eq:rotation}
\end{equation}
\par
As expected, the rotation (\ref{eq:rotation_ZW}) changes the sign of the topological charge density $\mathcal{W}$. Similarly, the sign of $\mathcal{Z}$ changes since the rotation swaps the red and blue triangles in Figure \ref{fig:ParSpaceTiles}. Finally, note that, as follows from (\ref{eq:rotation_XY}) and (\ref{eq:rotation_ZW}), the only hyperslope invariant with respect to the twelve-fold symmetry corresponds to the zero matrix. 

\subsection{Perpendicular area and average hyperslope}
\label{sec:patches}
For large patches of globally uniform square-triangle $\mathcal{ST}$ tilings, the following relation between the determinant of the average hyperslope $\mathcal{H}$ of a patch and its perpendicular area $\mathcal{A}_\bot$ holds asymptotically as the patch size approaches infinity \cite{imperor-clerc_square-triangle_2021}:
\begin{equation}
    \det(\mathcal{H}) = \frac{\mathcal{A}_{\bot}}{\mathcal{A}}
    \label{eq:det_relation}
\end{equation}
The proof of this result, as given in \cite{imperor-clerc_square-triangle_2021} for $\mathcal{ST}$ tilings, also holds for $\mathcal{STR}$ tilings, since the lifting procedure uses exactly the same 4D lattice. For finite patches, the equality (\ref{eq:det_relation}) is only approximately valid, with a small residual term that scales as the boundary-to-area ratio of the patch. However, there exists an important case of finite patches where (\ref{eq:det_relation}) is exact: a unit cell of a periodic tiling.
\par
Periodic $\mathcal{STR}$ tilings (often encountered as approximant phases, see Section \ref{sec:BaTiO}) are completely characterized by their unit cells. The two basis vectors defining a unit cell are integer linear combinations of vectors $\mathbf{e}_i$. Let $n$ and $m$ represent the tuples of integer coefficients in these combinations, and let 
$\mathbf{v}_n$ and $\mathbf{v}_m$ denote the corresponding basis vectors. These vectors can be lifted into the 4D space 
$\mathcal{P}\oplus \mathcal{P}_\bot$, yielding two non-collinear 4D vectors $\mathbf{V}_n$ and $\mathbf{V}_m$. Let $\mathcal{D}$ stand for the 2D plane spanned by these vectors. Then, as follows from (\ref{eq:plane_det}), we have:
$$
\det(\mathcal{H}_\mathcal{D})=\frac{\mathcal{A}_{\text{u.c.}\bot}}{\mathcal{A}_{\text{u.c.}}},
$$
where $\mathcal{H}_\mathcal{D}$ is the hyperslope of $\mathcal{D}$ and $\mathcal{A}_{\text{u.c.}}$ the unit cell area. It remains to show that $\mathcal{H}_\mathcal{D}$ equals the average hyperslope (\ref{eq:average_hyperslope}) of the unit cell considered as a finite patch of tiles. Note that $\mathcal{D}$ is a rational subspace of $\mathcal{P}\oplus \mathcal{P}_\bot$ (with respect to the lattice spanned by $(\bm{\epsilon}_i), i\in\{1,\dots,4\}$). In particular, it contains the 2D lattice spanned by  $\mathbf{V}_n$ and $\mathbf{V}_m$. Since all points of this lattice belong to the lifted tiling, the deviation of the latter from the plane $\mathcal{D}$ is bounded. Let us consider a large patch of periodic tiling composed of entire unit cells. The average hyperslope of such patch equals that of the unit cell. On the other hand, since the deviation of the lifted patch from $\mathcal{D}$ is bounded, it also equals $\mathcal{H}_\mathcal{D}$. Therefore, (\ref{eq:det_relation}) holds for a single unit cell of a periodic tiling.
\par 
For a general finite patch, equation (\ref{eq:det_relation}) does not hold, as illustrated by the example of a regular dodecagon. This shape can be filled, for instance, with twelve triangles and six squares, but also in many other ways (see Fig. \ref{fig:dodecagons}). However, as shown in Appendix \ref{sec:appendix_Geometry_4D}, the average hyperslope of a patch is completely defined by its boundary (see formula (\ref{eq:border_hyperslope})). Therefore, since the regular dodecagon has perfect twelve-fold symmetry, the average hyperslope of any patch of this shape is a zero matrix. On the other hand, for the regular dodecagon, we have $\mathcal{A}_\bot/\mathcal{A} =4\sqrt{3} -7 \neq 0$, which violates equation (\ref{eq:det_relation}).
\par
Another interesting example of a finite patch that does not tile the plane is the shield (Fig. \ref{fig:shield}(c)). Its average hyperslope is a purely antisymmetric matrix (\ref{eq:hyperslope_shield}), and its determinant $\det(\mathcal{H_{\text{shield}}})=1-\sqrt{3}/2$ also differs from the ratio $\mathcal{A}_\bot/\mathcal{A}=\sqrt{3}-2$ obtained from (\ref{eq:shield_areas}).
\par
In conclusion, relation (\ref{eq:det_relation}) is strictly valid only for a finite patch that can tile the plane in a periodic fashion or for an infinite globally uniform tiling like the above mentioned dodecagonal quasicrystals.
\section{Application to two-dimensional layers of Ba-Ti-O}
\label{sec:BaTiO}
\begin{figure*}
    \centering
    \includegraphics[width=\textwidth]{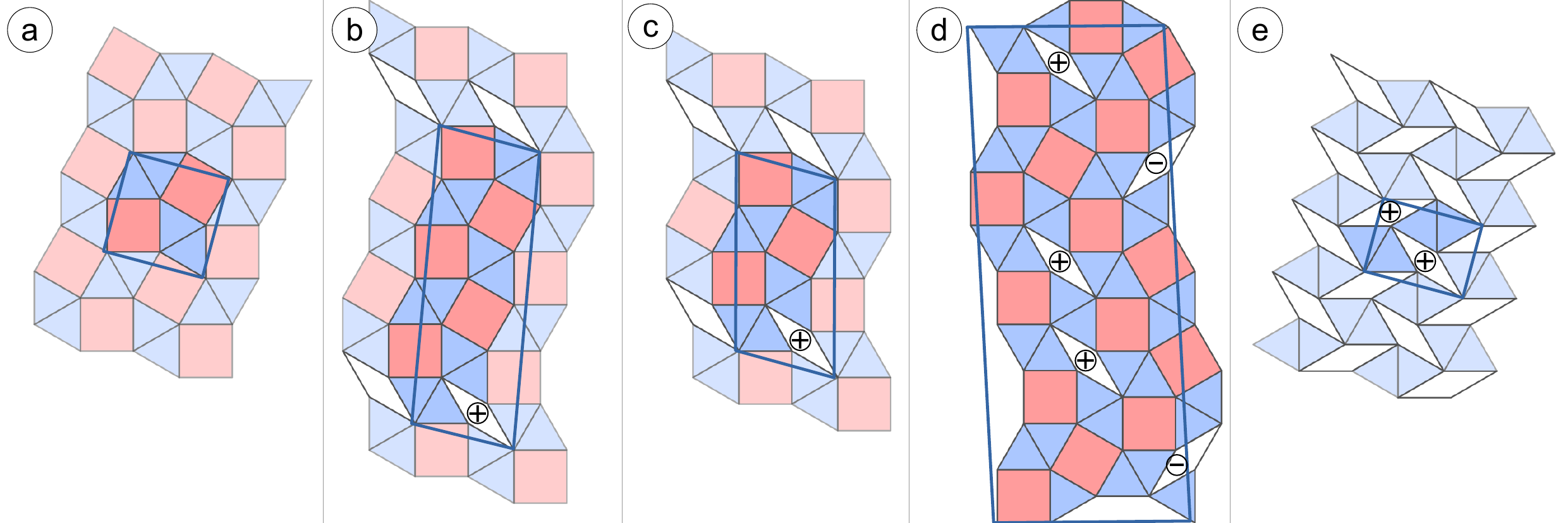}
     \caption{Series of approximant structures formed in Ba-Ti-O/Pd(111) for a Ba coverage of the Ti$_n$O$_n$ rings of (a)66.7\%, (b) 70.6\%, (c) 72.7 \%, (d) 72.7 \% and (e) 100 \%. The blue frames indicate the respective unit cells.}
      \label{fig:bto-aps}
\end{figure*}
In this section, we apply the higher-dimensional geometric approach introduced in section \ref{sec:tilings} to the characterization of dodecagonal oxide quasicrystals and approximant phases observed in two-dimensional oxides, such as Ba-Ti-O and Sr-Ti-O on a metal support \cite{forster_quasicrystalline_2013,schenk2017,forster2020}. In essence, these two-dimensional oxides are made of a network of Ti$_n$O$_n$ rings with n=4, 7, and 10 hosting Ba or Sr atoms \cite{schenk_2d_2022}. 
These alkaline earth metal atoms are visible in atomic-resolution scanning-probe images and form the vertices of $\mathcal{STR}$ tilings \cite{forster2020, wuehrl2022, wuehrl2023}. Oxide quasicrystals are commonly modeled by the ideal Niizeki-G\"ahler tiling (NGT) \cite{niizeki1987,gahler_crystallography_1988}. This tiling possesses a perfect twelve-fold symmetry, and therefore has zero average hyperslope. Thus, the overall area fractions of NGT satisfy the identity $\tau=\sigma + \rho=1/2$. The ratio of tiling elements $N_T:N_S:N_R$ in the NGT equals $(1+\sqrt{3}):1:(1+\sqrt{3})^{-1} \approx 2.73:1:0.37$ \cite{schenk_full_2019}, which corresponds to the following values in terms of area fractions for the infinite NGT: $\sigma=0.423$, $\rho=0.077$ and $\tau=0.5$.  The three different tiling elements occur with equal frequencies in all possible orientations. This results in a vanishing average topological charge. 
\par
Let us start by comparing the experimental data of a Ba-Ti-O OQC grown on Pt(111) \cite{schenk_full_2019} with the benchmark parameters of the NGT (Table \ref{tab:HyperslopeBTO}). This structure emerges when \SI{73.2}{\percent} of all Ti$_n$O$_n$ rings are occupied with Ba atoms. The analysis has been carried out for an ensemble of 8600 atomic vertices, which has been subject of a thorough statistical analysis before \cite{schenk_full_2019}. The outcome of the earlier study was that the overall ratio $N_T:N_S:N_R$ of 2.75 : 1 : 0.36 agrees well with the theorical figures for NGT. 
However, the analysis also reveals a significant breaking of the twelve-fold rotational symmetry. For instance, the occurrence of the squares in the orientation $\mathcal{S}_1$ exceeds that of $\mathcal{S}_2$ by roughly \SI{15}{\percent}, while the rhombuses in the orientations $\mathcal{R}_4$ and $\mathcal{R}_5$ are notably underrepresented. Most strikingly, dodecagonal clusters of the type shown in Figure \ref{fig:dodecagons}(b) were found almost exclusively in six out of twelve possible configurations. This fact could be explained by a local epitaxial stabilization of small patches of a complex approximant on the six-fold symmetric support. 
\par
Despite the discrepancies mentioned above, the overall area fractions for the OQC patch \cite{schenk_full_2019} agree remarkably well with the theoretical figures for NGT. For instance, in the case of squares the agreement is within three significant digits, and the area fraction of triangles exceeds that for NGT by only \SI{1.2}{\percent}. This small excess of triangles and the corresponding deficit of rhombuses supports the hypothesis of the breaking of the twelve-fold symmetry of the structure to the six-fold one. Note also that the value of the determinant of the average hyperslope $\det(\mathcal{H})$ of this OQC patch is very close to zero as well as its overall topological charge density $\mathcal{W}$. This analysis shows that this OQC patch has an average composition very close to a perfect NGT tiling. 
\begin{table}
	\begin{center}
		\caption{Comparison of different 2D STR tilings and the ideal dodecagonal Niizeki-G\"ahler tiling (NGT) regarding their Ba density $\rho_Ba$, their triangle, square and rhombus area coverage ($\tau, \sigma, \rho$), the determinant of the average hyperslope $\det(\mathcal{H})$, and their topological charge $\mathcal{W}$.\\}
		\label{tab:HyperslopeBTO}
		\begin{tabular}{|l|c|c|c|c|c|c|}
            \hline
            Phase & $\rho_{Ba}$ & $\tau$ & $\sigma$ & $\rho$ & $\det(\mathcal{H}$) & $\mathcal{W}$\\
           \hline
           \hline
            NGT &  & 0.500 & 0.423 & 0.077 & 0 & 0\\
            \hline
           \shortstack{OQC\\8600 vertices} & 73.2 \% & 0.506 & 0.423 & 0.071 & -0.00033 & -0.0049\\
           \hline
           \hline
           Fig. \ref{fig:bto-aps}(a) & 66.7 \% & 0.464 & 0.536 & 0 & -0.07180 & 0 \\
           \hline
           
           Fig. \ref{fig:bto-aps}(b) & 70.6 \% &0.486 & 0.467 & 0.047 & -0.0284 & 0.0810 \\
           \hline
           Fig. \ref{fig:bto-aps}(c)& 72.7 \% &0.497&0.431 & 0.072&-0.00515 & 0.1244 \\
           \hline
           Fig. \ref{fig:bto-aps}(d) & 72.7 \% &0.497 & 0.431 & 0.072 & -0.00515 & 0.0249\\
           \hline
           Fig. \ref{fig:bto-aps}(e) & 100 \% & 0.634 & 0 & 0.366 & 0.26795 & 0.6340\\
           \hline
           \hline
		\end{tabular}
	\end{center}
\end{table}
\par
It is instructive to compare the data for the OQC patch with that of its periodic approximants. When growing Ba-Ti-O layers on a Pd(111) substrate, a particularly interesting series of such approximants is observed \cite{wuehrl2022, wuehrl2023}. Starting at a Ti$_n$O$_n$ ring Ba coverage of 66.7\%, a periodic square-triangle tiling known as $\sigma$-phase forms (see Figure \ref{fig:bto-aps}(a)). Its small unit cell contains two squares and four triangles. When increasing the Ba coverage slightly, one-dimensional rows of rhombuses are introduced in the $\sigma$-phase tiling due to the formation of antiphase domain boundaries \cite{wuehrl2022}. The spacing of these boundaries is variable. The unit cell of the resulting $\mathcal{STR}$ tiling can be described by a combination of one rhombus, one square, and four triangles with multiple unit cells of the $\sigma$-phase. The two examples given in Fig. \ref{fig:bto-aps}(b,c) illustrate this combination with two (respectively one) $\sigma$-phase units. The highest domain-boundary density corresponds to a ring coverage of 72.7\%. However, at this level of coverage, an alternative periodic approximant structure shown in Fig. \ref{fig:bto-aps}(d) is also observed in experiments. Since its Ba coverage is very close to that for the dodecagonal quasicrystal formed on Pt(111), this structure should represent a close approximation of the quasicrystal. Finally, at a Ba coverage of 100\% of all Ti$_n$O$_n$ rings, a pure triangle-rhombus tiling shown in Fig. \ref{fig:bto-aps}(e)  emerges.
\par
As underlined previously, for approximant periodic phases, the determinant of the hyperslope $\det(\mathcal{H})$ gives a direct and quantitative measure of the difference from a perfect quasicrystal. Looking carefully to the benchmarks of these approximant structures listed in Tab. \ref{tab:HyperslopeBTO}, we note the following: The largest value of $\det(\mathcal{H})$, exceeding that of the OQC by orders of magnitude, is observed for the structure completely lacking squares (Figure \ref{fig:bto-aps}(e)). This can be explained by the fact that geometrical constraints exert a strong imbalance in the area fraction of triangles and that the topological charges of the same sign prevent from transforming rhombuses to squares. The lowest (and identical) absolute values of $\det(\mathcal{H})$ are found for the domain boundary patch consisting of eight triangles, three squares and one rhombus in Fig. \ref{fig:bto-aps}(c) and the most complex unit cell in Fig. \ref{fig:bto-aps}(d).  Both structures in Fig. \ref{fig:bto-aps}(c) and \ref{fig:bto-aps}(d) are indistinguishable in terms of area fractions. The value of $\det(\mathcal{H})$ is still a factor of 15 larger in comparison to that of OQC. These two structures demonstrate, that the complexity expressed in terms of the number of tiling elements in a unit cell does not provide by itself a sufficient measure for the degree to which a quasicrystal is approximated. It is also worth mentioning that the approximants of Figures \ref{fig:bto-aps}(c) and (d) have a different topological charge density, as can be seen from the value of the coefficient $\mathcal{W}$. This difference is due to the dilution effect. Indeed, despite the fact that the structure on Figure \ref{fig:bto-aps}(d) contains five rhombuses in the unit cell, the net topological charge of them equals $+1$. The same charge is found in the unit cell on Figure \ref{fig:bto-aps}(c), but the unit cell of the former structure is five times larger, and thus the coefficient $\mathcal{W}$ is five times smaller (see Table \ref{tab:HyperslopeBTO}). The same effect explains the 1.5 factor difference between the values of $\mathcal{W}$ for the structures on Figures  \ref{fig:bto-aps}(b) and (c). In contrast, the highest topological charge density is obtained for the small unit cell in Fig. \ref{fig:bto-aps}(e) containing two rhombuses of the same topological charge in a small unit cell. Thus, the data in Table \ref{tab:HyperslopeBTO} shows that the structural evolution in this sequence of approximant phases can be understood in terms of the coefficient $\mathcal{W}$ which is proportional to the topological charge density.   
\par
The most complex approximant shown in figure Fig. \ref{fig:bto-aps}(d) is a good example for testing the three different ways of calculating the hyperslope $\mathcal{H}$ introduced here: from the composition of tiles, from the border of the patch and directly from the two lattice vectors of the lifted unit cell.
This patch contains 40 triangles (10 in each of the four orientations $\mathcal{T}_1$, $\mathcal{T}_2$, $\mathcal{T}_3$, $\mathcal{T}_4$), 15 squares (10 $\mathcal{S}_1$, 3 $\mathcal{S}_2$ and 2 $\mathcal{S}_3$) and five rhombuses (2 $\mathcal{R}_2$ and 3 $\mathcal{R}_5$) and its global area fractions are $\tau=0.497$, $\sigma=0.431$ and $\rho=0.072$. According to (\ref{eq:area_ratio}), we arrive to a value of $\mathcal{A}_{\bot}/\mathcal{A}=-0.00515$. Alternatively, we can consider the area fractions of the individual prototiles for calculating the hyperslope according to (\ref{eq:average_hyperslope}). The resulting hyperslope is
$$
\mathcal{H}= \begin{pmatrix}
    0.0718 & - 0.0497 \\0 & -0.0718
\end{pmatrix}
$$ 
and its determinant equals $\det(\mathcal{H})=-0.00515$ as expected from (\ref{eq:det_relation}). 
When calculating first the four coefficients $\mathcal{X},\mathcal{Y},\mathcal{Z}$, and $\mathcal{W}$ using (\ref{eq:hyperslope_XYZW}), we get to the same result. 
\par
For calculating $\mathcal{H}$ from the border of the patch by applying expression (\ref{eq:border_hyperslope}), we need to define the path along the periphery in counterclockwise direction in steps of the four unit vectors. Starting from the bottom left corner, this path reads as: ($\mathbf{e}_1$, $\mathbf{e}_2$, $\mathbf{e}_1$, ($\mathbf{e}_2$-$\mathbf{e}_4$), $\mathbf{e}_4$, $\mathbf{e}_3$, $\mathbf{e}_4$, ($\mathbf{e}_3$-$\mathbf{e}_1$), $\mathbf{e}_4$, ($\mathbf{e}_3$-$\mathbf{e}_1$), $\mathbf{e}_4$, $\mathbf{e}_3$, $\mathbf{e}_4$, ($\mathbf{e}_3$-$\mathbf{e}_1$), ($\mathbf{e}_4$-$\mathbf{e}_2$), -$\mathbf{e}_1$, -$\mathbf{e}_2$, -$\mathbf{e}_1$, ($\mathbf{e}_1$-$\mathbf{e}_3$), -$\mathbf{e}_4$, -$\mathbf{e}_3$, -$\mathbf{e}_4$, ($\mathbf{e}_1$-$\mathbf{e}_3$), -$\mathbf{e}_4$, ($\mathbf{e}_1$-$\mathbf{e}_3$), -$\mathbf{e}_4$,-$\mathbf{e}_3$, -$\mathbf{e}_4$) and gives the same hyperslope matrix $\mathcal{H}$ as written above.
\par
Finally, the same result is obtained directly from the two lattice vectors of the unit cell. They correspond to the two tuples $n=(2,2,0,-1)$ and $m=(-3,0,5,5)$. The hyperslope $\mathcal{H}$ is directly obtained by applying (\ref{eq:hyperslope}) to the two 4D lifted unit cell vectors $\mathbf{V}_n$ and $\mathbf{V}_m$.

\section{Perspective}
\begin{figure}
    \centering
    \includegraphics[width=0.45\textwidth]{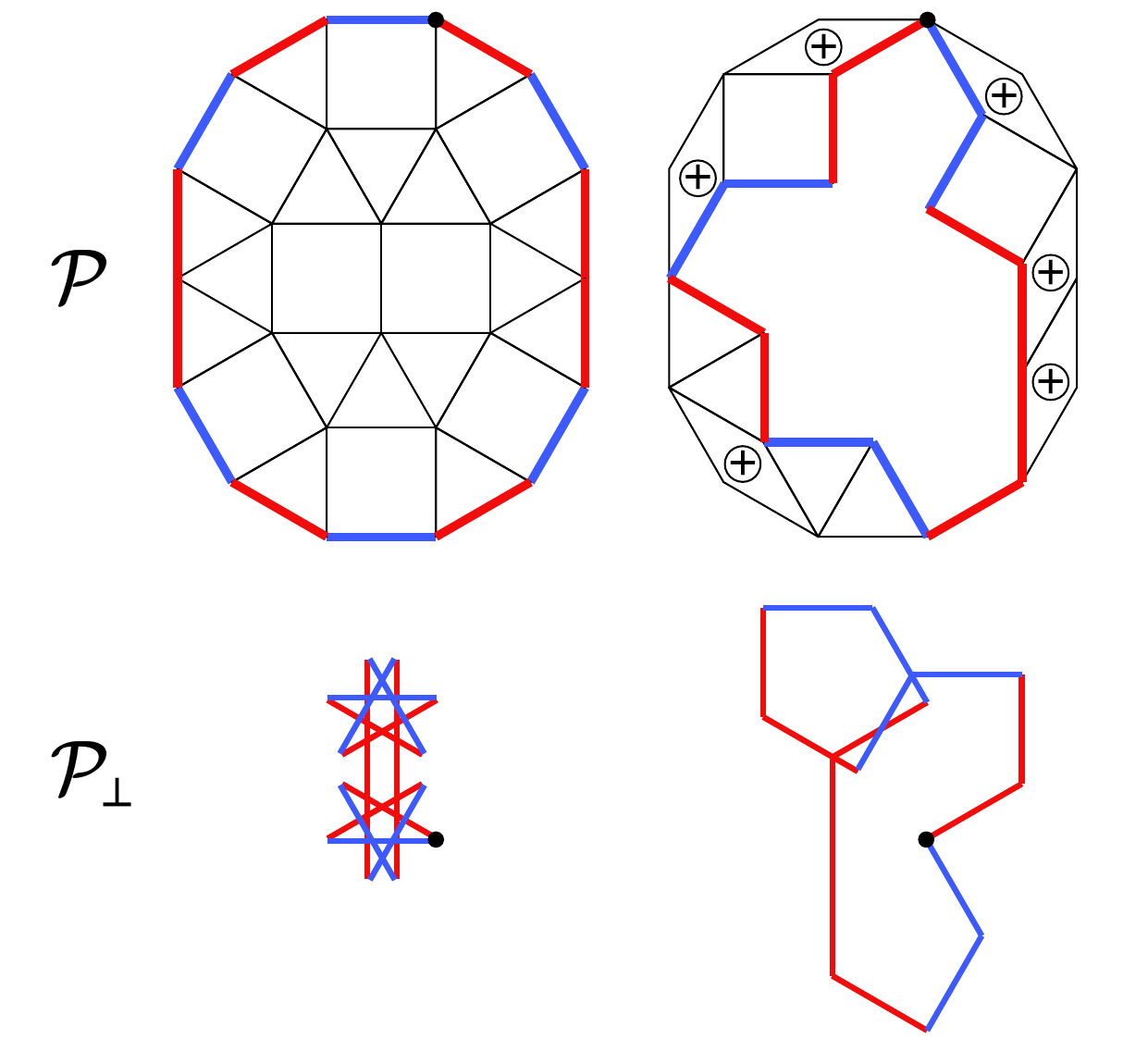}
     \caption{Chiral aperiodic monotile. From the $\mathcal{STR}$ patch in (a) with 14 edge vectors, the chiral aperiodic monotile (b) can be obtained by introducing tiles with negative algebraic areas in $\mathcal{P}$. The two projected borders in $\mathcal{P_{\bot}}$ are shown at the bottom.}
      \label{fig:monotile}
\end{figure}

A promising perspective of this work is to extend this approach to the recently discovered chiral aperiodic monotile \cite{smith2023chiral}. All 14 edges that form the boundary of this object are of equal length, and their directions are multiples of $\pi/6$. Although the monotile cannot be tiled with squares, triangles, or rhombuses, it can be represented as a formal difference between the two patches of $\mathcal{STR}$ tiling shown in Figure \ref{fig:monotile}, It is therefore possible to compute the formal average hyperslope of the monotile. This can be done either by averaging the hyperslopes of individual tiles as in (\ref{eq:average_hyperslope}) (with some tiles entering with a negative weight) or, alternatively, by directly lifting the boundary of the monotile into 4D space and using formula (\ref{eq:border_hyperslope}) for the average hyperslope. Both approaches yield the same result:
\begin{equation}
\begin{aligned}
    \mathcal{H}_{\text{mono}} &=\mathcal{X}_3 B^{S3}+ 
        \mathcal{Z} I_{2}+
        \mathcal{W} A^{R}/\sqrt{3} \\
        &=\frac{1}{12}\begin{pmatrix}
            -9+5\sqrt{3} & 9-3\sqrt{3} \\
            -27+9\sqrt{3} & -3 -\sqrt{3} 
        \end{pmatrix}\\ 
    \mathcal{X}_1 &= \mathcal{X}_2 =0\\
    \mathcal{X}_3 &= -\frac{1}{1+\sqrt{3}}\\
    \mathcal{Z} &= -\frac{1}{\sqrt{3}(1+\sqrt{3})}\\
    \mathcal{W} &= -\frac{\sqrt{3}}{(1+\sqrt{3})}
    \label{eq:monotile}
\end{aligned}
\end{equation}
for the orientation of the monotile shown in Fig. \ref{fig:monotile}. For the other orientations of the monotile, the average hyperslope  can be obtained by applying successive $\pi/6$ rotations and using formula (\ref{eq:rotation}). The area of the monotile is $\mathcal{A}_{\text{mono}}=3a^2(1+\sqrt{3})$. 
\par
The remarkable property of the monotile is that its shape enforces perfect matching rules for a quasiperiodic tiling \cite{smith2023chiral}. Notably, this tiling exhibits only six-fold rotational symmetry, despite each monotile having 12 possible orientations. Since rotation by $\pi/3$ conserves the topological charge of the monotile (which is -6 for the orientation shown in Figure \ref{fig:monotile}), one should expect a non-zero topological charge density in the quasiperiodic tiling by monotiles.

\section{Conclusion}

A general scheme for characterizing $\mathcal{STR}$ tilings based on lifting them into 4D space is proposed. The structure of an $\mathcal{STR}$ tiling in a bounded region is characterized by its average hyperslope, represented as a $(2\times 2)$ real matrix. This matrix is completely determined by the boundary of the patch. The average hyperslope imposes partial linear constraints on 13 area fractions, defining the composition of the tiling. The remaining 5 degrees of freedom (the local modes) correspond to local reconstructions of the tiling. These reconstructions involve the transformation of a square into a pair of rhombuses and vice versa. Such transformations can be understood as the creation or annihilation of a pair of topological charges of opposite sign.
\par
The average hyperslope of a tiling patch can be computed in two ways: either directly as an area-weighted average of the hyperslopes of individual tiles or indirectly from the boundary of the patch alone. This provides a quantitative method for characterizing the structure of real-world materials described by $\mathcal{STR}$ tilings, such as two-dimensional Ba-Ti-O layers.

\acknowledgments
Anuradha Jagannathan and Jean-François Sadoc (LPS, Orsay) are gratefully acknowledged for sharing ideas about quasiperiodic phases.

\bibliography{square-triangle-rhombus-April-2024.bib}

\appendix
\section{Geometry in 4D hyperspace}
\label{sec:appendix_Geometry_4D}
Useful geometrical relations for planes in the 4D hyperspace are given in this section. Note that all these relations are general and are not related to the 4D lattice. In the Euclidean space with four dimensions, a 4D orthonormal basis of four unit vectors $(\mathbf{I},\mathbf{J},\mathbf{K},\mathbf{L})$ can be introduced and any 4D-vector $\mathbf{X}$ has four coordinates in this basis:
\begin{equation*}
\mathbf{X}=x\mathbf{I}+y\mathbf{J}+x{_\bot}\mathbf{K}+y{_\bot}\mathbf{L}
    \label{eq:4Dcoordinates}
\end{equation*}
A (two-dimensional) plane in the 4D space is defined by two non-colinear 4D-vectors, as only planes containing the origin of space are considered here. Six different coordinate planes can be associated to the orthonormal basis: $(\mathbf{I},\mathbf{J})$, $(\mathbf{I},\mathbf{K})$, $(\mathbf{I},\mathbf{L})$, $(\mathbf{J},\mathbf{K})$, $(\mathbf{J},\mathbf{L})$ and $(\mathbf{K},\mathbf{L})$. For example, the plane $(\mathbf{I},\mathbf{J})$  contains the 4D-vectors of coordinates $(x, y, 0, 0)$. Among them, two planes, $(\mathbf{I},\mathbf{J})$ and $(\mathbf{K},\mathbf{L})$, are selected to play a special role. They are orthogonal to each other and any 4D-vector $\mathbf{X}$ can be represented by its two projections onto them using the two sets of coordinates $(x, y)$ and ($x{_\bot},y{_\bot})$. It is this projection scheme which is used in the lift construction of the physical plane $\mathcal{P}$ onto the 4D hyperspace. In a more formal way, the 4D Euclidean space can be defined as the orthogonal direct sum $\mathcal{P} \oplus \mathcal{P_\bot}$ where the planes $\mathcal{P}$ and $\mathcal{P_\bot}$ are two embedded 2D orthogonal subspaces. Note that very often, by extension, the plane $(\mathbf{I},\mathbf{J})$ in 4D is also called $\mathcal{P}$, even if it is a plane embedded in 4D hyperspace, with the same name as the physical plane $\mathcal{P}$.
\par
Let $\mathcal{D}$ be the plane spanned by two non-collinear 4D-vectors of coordinates $(a, b, c, d)$ and $(e, f, g, h)$ in the $(\mathbf{I},\mathbf{J},\mathbf{K},\mathbf{L})$ orthonormal basis. A vector $\mathbf{X} \in \mathcal{D}$ can thus be written in the following form:
\begin{equation*}
    \mathbf{X}=\begin{pmatrix}
        x \\ y \\ x{_\bot} \\ y{_\bot}
    \end{pmatrix} =\lambda
    \begin{pmatrix}
        a \\ b \\ c \\ d
    \end{pmatrix}+\mu
    \begin{pmatrix}
        e \\ f \\ g \\ h
    \end{pmatrix}
\end{equation*}
where $\lambda$ and $\mu$ are two real coefficients. Whenever $af-be \neq 0$, the components of $\mathbf{X}$ satisfy the following relation:
\begin{equation*}
\begin{pmatrix}
    x{_\bot} \\ y{_\bot} 
\end{pmatrix} = \mathcal{H}_\mathcal{D}
\begin{pmatrix}
    x \\ y 
\end{pmatrix}    
\end{equation*}
where the $2\times 2$ matrix $\mathcal{H}_\mathcal{D}$ is given by the formula
\begin{equation}
\mathcal{H}_\mathcal{D} =\frac{1}{(af-be)}
\begin{pmatrix}
    (cf-bg) & (ag-ce) \\(df-bh) & (ah-de)
\end{pmatrix}  
\label{eq:hyperslope}
\end{equation}
The matrix $\mathcal{H}_\mathcal{D}$ is called the hyperslope of $\mathcal{D}$ relative to the plane $\mathcal{P}$ (it is important to underline here that a hyperslope is always relative to a reference plane). Note that $af-be=0$ if and only if $\mathcal{D}=\mathcal{P}_\bot$, which would correspond to an infinite hyperslope.
\par
It is worth noting that the terms in the formula (\ref{eq:hyperslope}) can be given a clear geometrical interpretation. Indeed, we can recognize in these terms the formulas for the determinant, or equivalently the signed area of a parallelogram. For instance, $(af-be)$ is the signed area of the parallelogram in the plane $\mathcal{P}$ spanned by the two vectors $(a,b)$ and $(e,f)$. Similarly, the other terms can be interpreted as signed areas of the parallelograms in the planes $(\mathbf{J},\mathbf{K})$, $(\mathbf{I},\mathbf{K})$, $(\mathbf{J},\mathbf{L})$ and $(\mathbf{I},\mathbf{L})$. Note that all these parallelograms are projections of the same parallelogram $\Pi$ in 4D space, spanned by the vectors $(a, b, c, d)$ and $(e, f, g, h)$, on the corresponding coordinate planes. Thus the hyperslope (\ref{eq:hyperslope}) can be written as
\begin{equation}
\mathcal{H}_\mathcal{D} =\frac{1}{\mathcal{A}_{(x,y)}}\begin{pmatrix}
    -\mathcal{A}_{(y,x_{\bot})} & \mathcal{A}_{(x,x_{\bot})} \\
    -\mathcal{A}_{(y,y_{\bot})} & \mathcal{A}_{(x,y_{\bot})}
\end{pmatrix},  
\label{eq:plane_hyperslope}
\end{equation}
where $\mathcal{A}_{(x,y)}$ is the signed area of $\Pi$ projected on $\mathcal{P}$, with the similar notations used for other projections. Note that any couple of non-colinear vectors $(a, b, c, d)$ and $(e, f, g, h)$ in $\mathcal{D}$ will give the same hyperslope.
\par
Finally, a useful relation is given by the value of the determinant $\det(\mathcal{H}_\mathcal{D})$, as we recognize the quantity $(ch-dg)=\mathcal{A}_{(x_{\bot},y_{\bot})}$, the projected area of the 4D parallelogram $\Pi$ over the plane $(\mathbf{K},\mathbf{L})=\mathcal{P_{\bot}}$: 
\begin{equation}
    \det(\mathcal{H}_\mathcal{D})=\frac{ch-dg}{af-be}=\frac{\mathcal{A}_{(x_{\bot},y_{\bot})}}{\mathcal{A}_{(x,y)}}
    \label{eq:plane_det}
\end{equation}
In conclusion, we see that a 4D parallelogram of the hyperspace is giving rise to six different signed areas.
\par
When considering the average hyperslope $\mathcal{H}$ of any lifted patch of tiles, we can combine the above expression of the hyperslope $\mathcal{H}_\mathcal{D}$ for a single 4D parallelogram with the Surveyor's area formula \cite{Surveyor_Braden_1986}. The later gives the oriented area $\mathcal{A}_{tot}$ of any polygon in the plane from the coordinates of its vertices. These vertices $((x_0,y_0), (x_1,y_1), ... (x_{n-1},y_{n-1}))$ are counted counterclockwise along the border of the polygon and the Surveyor's area formula reads:
\begin{equation*}
\mathcal{A}_{\text{tot}}=\frac{1}{2}((x_0 y_1 - y_0 x_1)+ ... + (x_{n-1} y_0 - y_{n-1} x_0))
\end{equation*}
The average hyperslope of a patch (\ref{eq:average_hyperslope}) is obtained by averaging the hyperslopes of its tiles weighed with their areas. Taking into account (\ref{eq:plane_hyperslope}), this yields
\begin{equation}
\mathcal{H} =\frac{1}{\mathcal{A}_{\text{tot}(x,y)}}\begin{pmatrix}
    -\mathcal{A}_{\text{tot}(y,x_{\bot})} & \mathcal{A}_{\text{tot}(x,x_{\bot})} \\
    -\mathcal{A}_{\text{tot}(y,y_{\bot})} & \mathcal{A}_{\text{tot}(x,y_{\bot})}
\end{pmatrix},  
\label{eq:border_hyperslope}
\end{equation}
where the matrix elements are obtained by applying the Surveyor's area formula to the border of the patch projected onto the corresponding coordinate planes.
When we consider the determinant of the hyperslope, we get a more general relation: 
\begin{equation}
    \det(\mathcal{H})=\frac{\mathcal{A}_{\text{tot}(x,x_{\bot})}\mathcal{A}_{\text{tot}(y,y_{\bot})}-\mathcal{A}_{\text{tot}(x,y_{\bot})}\mathcal{A}_{\text{tot}(y,x_{\bot})}}{{\mathcal{A}^2_{\text{tot}(x,y)}}}
    \label{eq:border_hyperslope_det}
\end{equation}

\end{document}